\documentclass{article}

\usepackage{arxiv}

\usepackage[utf8]{inputenc} % allow utf-8 input
\usepackage[T1]{fontenc}    % use 8-bit T1 fonts
\usepackage{hyperref}       % hyperlinks
\usepackage{url}            % simple URL typesetting
\usepackage{booktabs}       % professional-quality tables
\usepackage{amsfonts}       % blackboard math symbols
\usepackage{nicefrac}       % compact symbols for 1/2, etc.
\usepackage{microtype}      % microtypography
\usepackage{lipsum}		% Can be removed after putting your text content
\usepackage{graphicx}
\usepackage{amsmath,amssymb}
\usepackage{caption}
\usepackage{subcaption}
\usepackage{natbib}
\usepackage{doi}
\usepackage{booktabs}

\title{High-Visibility Franson Interference Enabled by Passive Photonic Integrated Interferometers at Telecom Wavelengths}

\author{
\href{https://orcid.org/0000-0003-3576-2286}{\includegraphics[scale=0.06]{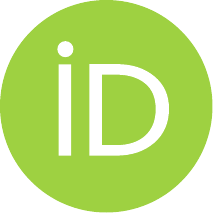}\hspace{1mm}Ramin Emadi}\thanks{Corresponding author.} \\
CNR--INO, Largo Enrico Fermi 6, Firenze, Italy \\
\And
\href{https://orcid.org/0000-0001-5249-8339}{\includegraphics[scale=0.06]{orcid.pdf}\hspace{1mm}Domenico Ribezzo} \\
CNR--INO, Largo Enrico Fermi 6, Firenze, Italy \\
Dipartimento di Fisica, Università degli Studi di Firenze, via G.~Sansone~1, Sesto Fiorentino, Italy \\
\And
\href{https://orcid.org/0009-0008-4194-9251}{\includegraphics[scale=0.06]{orcid.pdf}\hspace{1mm}Giulia Guarda} \\
CNR--INO, Largo Enrico Fermi 6, Firenze, Italy \\
European Laboratory for Non-Linear Spectroscopy (LENS), Sesto Fiorentino, Italy \\
\And
\href{https://orcid.org/0000-0002-7757-4331}{\includegraphics[scale=0.06]{orcid.pdf}\hspace{1mm}Davide Bacco} \\
CNR--INO, Largo Enrico Fermi 6, Firenze, Italy \\
Dipartimento di Fisica, Università degli Studi di Firenze, via G.~Sansone~1, Sesto Fiorentino, Italy \\
QTI srl, Largo Enrico Fermi 6, Firenze, Italy \\
\And
\href{https://orcid.org/0000-0002-1359-7956}{\includegraphics[scale=0.06]{orcid.pdf}\hspace{1mm}Alessandro Zavatta} \\
CNR--INO, Largo Enrico Fermi 6, Firenze, Italy \\
QTI srl, Largo Enrico Fermi 6, Firenze, Italy \\
}

\renewcommand{\shorttitle}{High-Visibility Franson Interference...}

\hypersetup{
pdftitle={A template for the arxiv style},
pdfsubject={q-bio.NC, q-bio.QM},
pdfauthor={David S.~Hippocampus, Elias D.~Striatum},
pdfkeywords={First keyword, Second keyword, More},
}

\begin{document}
\maketitle

\begin{abstract}
	High-visibility Franson interference at telecom C-band wavelengths is achieved using a cascaded periodically poled lithium niobate (PPLN) waveguide photon-pair source combined with fully passive, path-imbalanced Mach–Zehnder interferometers implemented on photonic integrated circuits (PICs). The interferometers require neither on-chip phase shifters nor active stabilization; instead, the phase is scanned via thermal tuning of the chip. By employing a narrow-linewidth continuous-wave (CW) pump and dense wavelength-division multiplexing (DWDM) filtering, energy-time entangled photon pairs with high spectral indistinguishability are generated. We achieve a 4.8\% heralding efficiency and a two-photon
interference visibility of 97.1\% from sinusoidal fringe fitting (raw visibility 95.2\% and
background-corrected visibility 95.6\%), alongside a coincidence-to-accidental ratio (CAR)
exceeding $10^{3}$ at only 1.7~mW of pump power. These results represent one of the highest
Franson-interference visibilities reported for a PIC-based analyzer within a compact,
fiber-integrated platform.
\end{abstract}

% keywords can be removed

\section{Introduction}
Long-distance distribution of entangled states is a fundamental requirement for
interconnecting emerging quantum technologies in a quantum internet~\cite{paudel2023quantum,kimble2008quantum,wehner2018quantum}.
Meeting this challenge requires entanglement sources and single-photon detectors
that are compatible with existing telecom-fiber infrastructure~\cite{laucht2021roadmap,politi2009integrated}.
Entanglement can be generated in several degrees of freedom, including polarization,
spatial modes (including orbital angular momentum and few-mode fibers), frequency bins,
and energy--time. Among these, energy--time entanglement is particularly attractive for
fiber networks because it is robust to polarization fluctuations and compatible with dense
wavelength-division multiplexing (DWDM), as demonstrated in long-distance field
trials~\cite{marcikic2004distribution,tittel2000quantum,gisin2007quantum}. Energy–time entanglement offers intrinsic robustness in standard telecom fiber and seamless compatibility with DWDM, unlike polarization, spatial-mode, or frequency-bin encodings that require active compensation or stringent phase control~\cite{tittel2000quantum,marcikic2004distribution,xavier2025energy}.

A common method to analyze energy-time entangled photon pairs uses two unbalanced
Mach-Zehnder interferometers (uMZIs) in the Franson configuration~\cite{franson1989bell,marcikic2004distribution}.
Each photon from a spontaneous parametric down-conversion (SPDC) pair enters an
interferometer and can travel along a \textit{short} (S) or \textit{long} (L) path, leading to four path
combinations: $S_1S_2$, $S_1L_2$, $L_1S_2$, and $L_1L_2$. Under continuous-wave
pumping, and when the pump coherence time is much longer than the interferometers'
imbalance, the emission time is uncertain enough that the $S_1S_2$ and $L_1L_2$
contributions are indistinguishable in the coincidence-time difference. As a result,
two-photon interference occurs when the single-photon coherence time is much
shorter than the path delay, suppressing single-photon interference~\cite{franson1989bell,marcikic2004distribution}. This effect---Franson interference---appears as a phase-dependent modulation of the
coincidence rate as a function of the relative phase between the interferometers~\cite{marcikic2004distribution}.

%------------Missing Paragraph

Early demonstrations of Franson interference in fiber-based SPDC systems confirmed the
nonclassical nature of these correlations, but also highlighted practical challenges, including
precise matching of interferometer delays, long-term phase stability, and suppression of
accidental coincidences (e.g., due to dark counts and multi-pair emission)~\cite{rarity1990experimental,brendel1991time,marcikic2004distribution}.
High-visibility two-photon interference has also been demonstrated in picosecond-regime separated-source
architectures at telecom wavelengths~\cite{aboussouan2010high},
representing a complementary approach to
continuous-wave (CW) and narrowband-engineered guided-wave sources.
Moreover, standard Franson interferometry relies on post-selection: only the indistinguishable
$S_1S_2$ and $L_1L_2$ events contribute to the interference, whereas the distinguishable
$S_1L_2$ and $L_1S_2$ events are discarded.
Consequently, the basic Franson analyzer is incompatible with device-independent security protocols unless
supplemented by additional measures such as active switching or true time-bin schemes~\cite{bacco2019boosting}.
For many network applications that do not require device-independent security, however, Franson interference
remains a practical and informative diagnostic of energy--time entanglement quality.

Recent advances in integrated photonics and engineered nonlinear media are enabling
scalable entangled-photon sources for deployment in fiber networks~\cite{wang2020integrated,boyd2008nonlinear}.
Periodically poled lithium niobate (PPLN) waveguides are particularly attractive because
quasi-phase matching, strong confinement, and thermo-optic tunability enable bright and
stable photon-pair generation with efficient fiber coupling~\cite{rarity1990experimental}. In parallel, photonic
integrated circuits (PICs) provide compact uMZI analyzers with strong common-mode
rejection of environmental perturbations, reducing the need for continual active phase
locking~\cite{halder2008high,zhang2021high}. For example, entanglement distributed over multicore fiber has been analyzed
using on-chip interferometers while maintaining visibility above 94\% without active
tuning~\cite{da2021path}. In parallel, guided-wave and fully fibered architectures continue to provide highly deployable entanglement sources compatible with DWDM networking.
For instance, Troisi \textit{et al.} reported a fully fibered nonlinear Sagnac source based on cascaded PPLN waveguides, supporting broadband DWDM multiplexing
and demonstrating raw Franson visibilities of $99\pm1\%$ on a standard 100~GHz ITU channel pair (19,23) with a passively stabilized fiber analyzer~\cite{Troisi2026Sagnac}.

In this work, we present a compact, fiber-pigtailed source-analyzer platform that combines a
cascaded PPLN waveguide photon-pair source with fully passive PIC-based Franson
interferometers. The cascaded SHG$\to$SPDC architecture provides a narrowband pump for
SPDC and improves spectral indistinguishability, while the monolithic PIC uMZIs provide robust
phase stability without on-chip heaters or electro-optic phase shifters. We observe a raw
two-photon interference visibility of 95.2\% (95.6\% after accidental subtraction) and a
fit-derived visibility of 97.1\% at 1550~nm, together with a coincidence-to-accidental ratio
(CAR) exceeding $10^{3}$ at milliwatt-level pump power. The platform is fully fiber integrated and
operates on standard DWDM channels around 1560~nm, underscoring its suitability for
telecom-compatible quantum networking.

\begin{figure}[t]
  \centering
  \includegraphics[width=\linewidth]{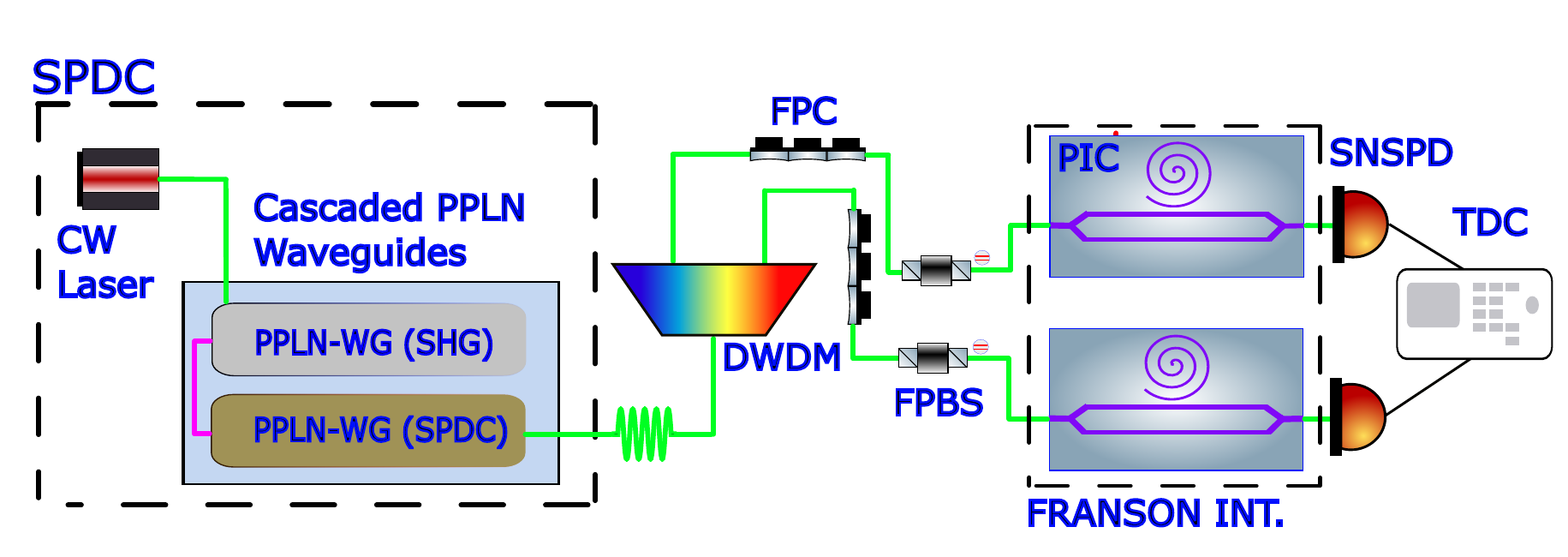}
  \caption{Experimental setup. 
A CW laser at 1560~nm pumps two cascaded 
periodically poled lithium niobate (PPLN) waveguides 
(second-harmonic generation, SHG, followed by spontaneous parametric down-conversion, SPDC). 
A DWDM separates the signal (CH22) and idler (CH20), 
each entering a matched unbalanced Mach--Zehnder interferometer (uMZI) 
with a path difference of approximately $\Delta t \approx 0.8$~ns after polarization cleaning. 
Detection is performed using superconducting nanowire single-photon detectors (SNSPDs) 
and a time-to-digital converter (TDC) for coincidence counting and visibility analysis.}
  \label{fig:setup}
\end{figure}

\section{Materials and Methods for Energy–Time Entanglement Generation}

\subsection{Cascaded PPLN Source Design and Experimental Setup}

\begin{figure}[t]
\centering

% ---------- Left image ----------
\begin{subfigure}[t]{0.5\textwidth}
    \centering
    \includegraphics[width=\linewidth, height=6.35cm, keepaspectratio]{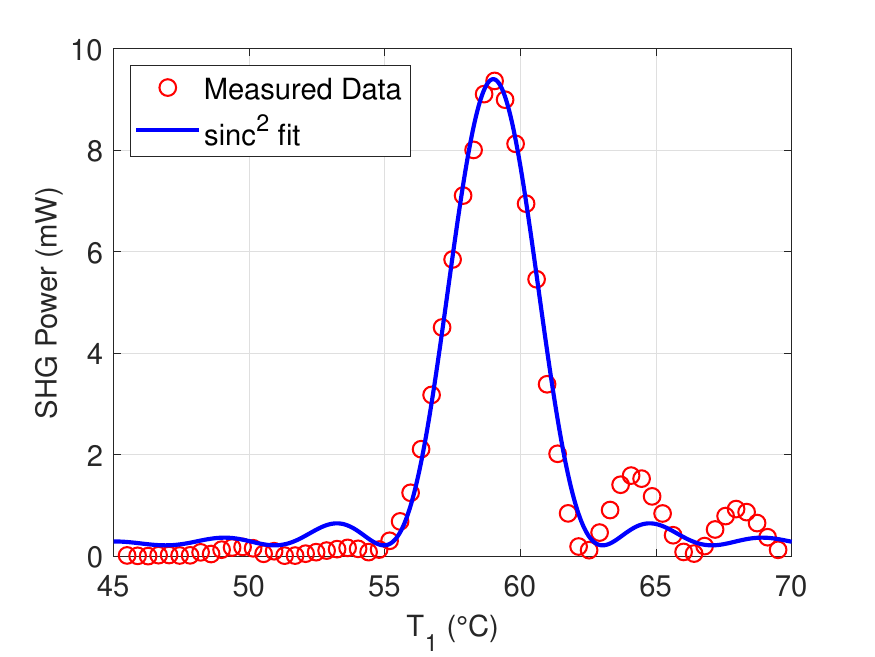}
    \caption{} % This automatically adds "(a)"
    \label{fig:shg_tuning}
\end{subfigure}
\hfill
% ---------- Right image ----------
\begin{subfigure}[t]{0.45\textwidth}
    \centering
    \includegraphics[width=\linewidth, height=6cm, keepaspectratio]{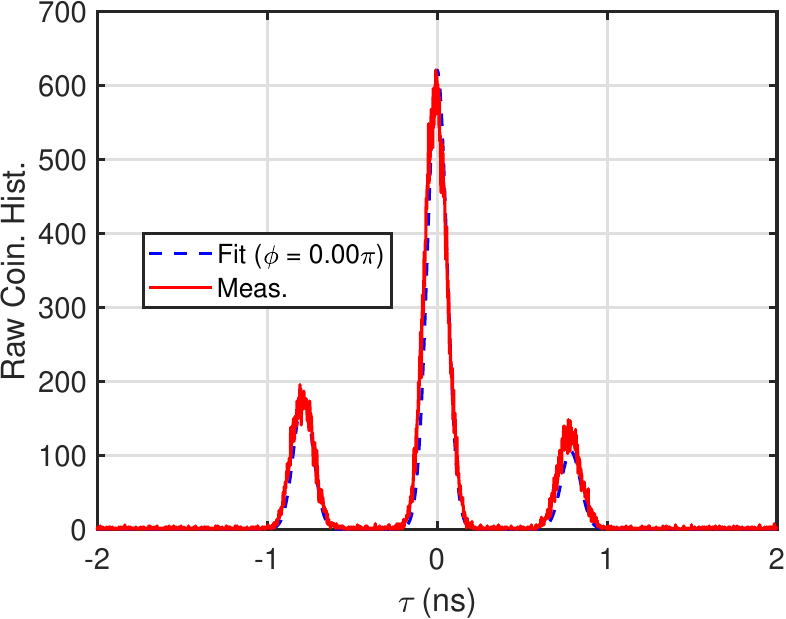}
    \caption{} % This automatically adds "(b)"
    \label{fig:pair_rate_vs_power}
\end{subfigure}

\caption{(a) Temperature tuning of the PPLN waveguide measured via SHG. 
Open circles: measured SHG power; solid curve: $\mathrm{sinc}^{2}$ fit. 
The extracted FWHM of $\approx 3.53^{\circ}\mathrm{C}$ gives the 
temperature-acceptance bandwidth of the QPM structure. The peak temperature sets the operating point for SPDC. 
(b) Measured coincidence histogram (red) together with a fit to the 
two-photon correlation model of Eq.~\eqref{eq:franson_hist_model} for the constructive 
interference setting $\phi = 0$ (blue dashed). The three Gaussian peaks correspond to the LS ($\tau \approx -\Delta t$),
SS/LL ($\tau \approx 0$), and SL ($\tau \approx +\Delta t$) path contributions of the
Franson interferometer, with only the central SS/LL peak exhibiting phase-dependent
modulation.}
\label{fig:g2_fit}
\end{figure}

A schematic of the experimental setup is shown in Fig.~\ref{fig:setup}. The entanglement source begins with a narrow-linewidth CW fiber laser at 1560.48~nm (telecom C-band), which serves as the pump. The laser (linewidth $\Delta\nu_{p} \approx 7.6$~kHz) delivers up to 2~W of optical power to drive the nonlinear cascade. The first stage is a PPLN waveguide module operated as a SHG: it converts the 1560~nm pump light to 780~nm (second harmonic) via a quasi-phase-matched $\chi^{(2)}$ interaction. The waveguide is implemented in magnesium-oxide-doped PPLN with periodic poling and is fiber-coupled at both input and output. Temperature tuning of this waveguide controls the phase-matching condition for SHG. By adjusting the crystal temperature, we can scan the SHG efficiency curve, which follows the expected $\mathrm{sinc}^{2}$ dependence on phase mismatch~\cite{fejer2002quasi,myers1995quasi}. Figure~\ref{fig:g2_fit}(a) shows the measured SHG output power versus waveguide temperature for the first PPLN stage, displaying the characteristic $\mathrm{sinc}^{2}$ lobes of the QPM response. A fit to the main lobe yields a temperature full width at half maximum (FWHM) of approximately 3.5~$^{\circ}$C~\cite{krapick2013efficient}, consistent with typical acceptance bandwidths (2--5~$^{\circ}$C) for type-0 PPLN waveguides of similar length~\cite{spring2013chip}. This indicates that the SHG stage provides a relatively narrowband pump centered at 780~nm, which is beneficial for generating spectrally indistinguishable photon pairs in the next stage.

The second stage of the cascade is a PPLN waveguide (of the same or similar device family) operated in SPDC mode to produce signal and idler photons around 1560~nm. We use a type-0 phase-matching configuration (all interacting waves polarized along the extraordinary axis of the crystal) to maximize the nonlinearity (exploiting the largest tensor element $d_{33}$)~\cite{boyd2008nonlinear,fejer2002quasi}. The 780~nm light from the SHG stage (the pump for SPDC) is coupled into this waveguide to drive parametric down-conversion. Energy conservation dictates that each pump photon at frequency $\omega_{p}$ splits into signal ($\omega_{s}$) and idler ($\omega_{i}$) photons such that $\omega_{p} = \omega_{s} + \omega_{i}$~\cite{boyd2008nonlinear,hong1986experimental}. The SPDC phasematching in a uniform waveguide also yields a sinc-shaped spectral amplitude function: the probability amplitude for pair generation is proportional to the integral of the nonlinear interaction over the length $L$ of the waveguide, $\int_{0}^{L} \exp(i\Delta k z)\,dz \propto \mathrm{sinc}(\Delta k L/2)$, where $\Delta k$ is the wavevector mismatch at the given temperature~\cite{fejer2002quasi}. Thus, the SPDC joint spectral intensity has a main lobe width inversely related to the waveguide length, analogous to the SHG phase-matching envelope~\cite{tanzilli2001highly,mosley2008heralded}. In practice, a relatively broad phasematching bandwidth (several nanometers) is typical for type-0 SPDC in PPLN, meaning the generated photons are initially emitted with a range of frequencies. We employ external filtering to define the final photon-pair wavelengths and bandwidths, as described next.

After the cascaded PPLN module, the output (around 1560~nm, containing both signal and idler photons plus residual pump) is routed through a 32-channel DWDM demultiplexer, which separates wavelengths on the standard ITU grid. We select two specific DWDM channels to pass the signal and idler: channel~22 (center $\sim$1559.8~nm) and channel~20 ($\sim$1561.4~nm)~\cite{harder2013optimized,zhang2021high}. These channels are 200~GHz apart in frequency (two 100-GHz-spaced channels, corresponding to about 1.6~nm around 1560~nm), roughly centered around the degeneracy point (1560.6~nm) of the SPDC process. The DWDM filter has a 100~GHz passband for each channel and insertion loss $<5.4$~dB. This filtering accomplishes two important tasks: (1) it defines the signal and idler wavelengths, eliminating spectral overlap or ambiguity, and (2) it sets the single-photon coherence time $\tau_{c}$ by cutting down the SPDC bandwidth. For a filter of bandwidth $\Delta\nu$ (in Hz), the coherence time is approximately $\tau_{c} \approx 1/(\pi \Delta\nu)$~\cite{mandel1995optical,goodman2015statistical}. With a 100~GHz DWDM filter, the single-photon coherence time is on the order of a few
picoseconds ($\sim 3$~ps), depending on the exact filter shape~\cite{mandel1995optical,goodman2015statistical}.

With a 100~GHz ($\sim$0.8~nm) DWDM filter, we obtain $\tau_{c}$ on the order of $\sim 3$~ps. This short coherence time is much smaller than the path delay ($\sim$0.8~ns) in our interferometers, thereby suppressing single-photon interference while still allowing two-photon time--energy entanglement to be observed (via the Franson scheme).

Each filtered photon (signal and idler) is coupled into an uMZI on a photonic chip for analysis. The interferometers are fabricated on a thermally stable borosilicate glass platform, with evanescently coupled waveguide splitters/combiners~\cite{guarda2024quantum}. The path-length difference in each uMZI corresponds to a temporal delay $\Delta t \approx 0.8$~ns between the short and long arms~\cite{guarda2024quantum}. This delay is intentionally much larger than the single-photon coherence time ($\sim 3$~ps), so that any single-photon interference is averaged out, while entangled photon pairs can still produce two-photon Franson interference (since both photons either take the short paths or the long paths, arriving coincidentally despite the delay). A key feature of these PIC interferometers is that they contain no active phase shifters or heaters on chip. They are purely passive devices, meaning the relative phase $\phi$ between the two arms is not tuned by, say, microheaters (which would require electrical power and feedback control). Instead, we adjust $\phi$ globally by controlling the chip temperature: a uniform temperature change alters the effective refractive index of the glass waveguides via the thermo-optic effect, thus shifting the interferometric phase. This approach yields excellent phase stability and repeatability without any electronic stabilization, at the cost of slower tuning speed (thermal tuning has a response on the order of seconds, but this is sufficient for static or slowly varying phase scans in an experiment). The interferometers were packaged such that we could mount the PICs on thermoelectric heaters for coarse tuning; in practice, once set, the phase drifted only minimally (owing to the common-mode stability of the monolithic interferometer design). Each PIC interferometer exhibited a total fiber-to-fiber insertion loss of 6.9~dB at 1560~nm (including the intrinsic 3~dB splitting loss), i.e., approximately 1--1.5~dB higher than the nominal characterization value near 1545~nm, consistent with off-design operation and polarization-dependent coupling. The pair of PIC interferometers maintained a stable phase relationship over hours of operation, with feedback only needed for deliberate scans.

Finally, after passing through the interferometers, the photons are detected by fiber-coupled single-photon detectors. In our setup, we used superconducting nanowire single-photon detectors (SNSPDs) for both signal and idler, providing high detection efficiency and low timing jitter ($\sim$50~ps). A time-correlated single-photon counting system was used to record detection timestamps and construct coincidence histograms.

\subsection{Measurement Procedures and Data Analysis}
\label{Meas&Anlss}

We characterize the system in terms of source brightness, heralding efficiency, noise (accidental coincidences), and two-photon interference visibility. The detected singles count rates for signal ($S_s$) and idler ($S_i$), as well as the coincidence count rate $C$ (within a specified time window), are measured as functions of the pump power. Because the detectors have finite dead time and dark counts, we apply several corrections and definitions as follows.

The raw singles counts are first background-subtracted by measuring the detector dark count rates ($D_s$, $D_i$) and subtracting these from $S_s$ and $S_i$. We then correct for detector dead time (50~ns per count in our SNSPD system) using a non-paralyzable model~\cite{knoll2010radiation,harder2013optimized}, which yields effective singles rates $S'_s$ and $S'_i$:

\begin{equation}
S'_s = \frac{S_s - D_s}{1 - (S_s - D_s)\tau_{\mathrm{dead}}}, \qquad
S'_i = \frac{S_i - D_i}{1 - (S_i - D_i)\tau_{\mathrm{dead}}}
    \label{Eq:SingletRates}
\end{equation}

with $\tau_{\mathrm{dead}} \approx 50$~ns for each detector. This correction accounts for the slight undercounting at high rates due to the detectors being briefly inactive after each detection event. Next, we estimate the accidental coincidence rate (background coincidences) expected given the measured singles. In a continuous-wave-pumped SPDC source, true coincidences (from paired photons) scale with the pair generation rate, whereas accidentals arise from coincidences between uncorrelated detections (e.g., photons from different pair events or dark counts). For low pair generation probability, a Poisson model can be used in which the accidental rate $A_h$ is approximated by the product of the singles rates and the coincidence timing window $\tau_w$~\cite{aboussouan2010high,krapick2013efficient}. We take $\tau_w = 200$~ps, corresponding to our coincidence histogram bin width. Thus,
\begin{equation}
A_h \approx S'_s S'_i \tau_w
    \label{Eq:AccidentialRateEstm}
\end{equation}
which gives the accidental count rate expected from overlapping two independent photon streams with rates $S'_s$ and $S'_i$. We also verify accidentals by inspecting the flat baseline of the measured coincidence histogram away from the main peaks~\cite{aboussouan2010high}. From the measured total coincidence count rate $C_{\mathrm{meas}}$ (integrated around the zero-delay peak of the histogram), we obtain the true coincidence rate (coincident pairs) by subtracting accidentals~\cite{aboussouan2010high,harder2013optimized}:
\begin{equation}
C_{\mathrm{true}} = C_{\mathrm{meas}} - A_h
    \label{Eq:TrueConicEstimation}
\end{equation}

Using these quantities, we calculate the heralding efficiency for each channel, which quantifies the probability that detecting one photon (e.g., idler) heralds the presence of its entangled partner (signal). The signal heralding efficiency is~\cite{klyshko1980use,aboussouan2010high}
\begin{equation}
\eta_s = \frac{C_{\mathrm{true}}}{S'_i}
    \label{Eq:HeraldingEff_Signal}
\end{equation}

and similarly the idler heralding efficiency is
\begin{equation}
\eta_i = \frac{C_{\mathrm{true}}}{S'_s}
    \label{Eq:HeraldingEff_Idler}
\end{equation}

These represent system efficiencies from pair creation to detection for each arm and are influenced by losses (e.g., fiber coupling, filtering, detector efficiency) in the signal or idler paths. We also infer the internal pair generation rate $R_{\mathrm{pair}}$ (i.e., pairs generated in the crystal prior to losses) using the relation~\cite{aboussouan2010high,krapick2013efficient}
\begin{equation}
R_{\mathrm{pair}} = \frac{C_{\mathrm{true}}}{\eta_s \eta_i}
    \label{Eq:PairGenRate}
\end{equation}

This effectively divides the true coincidence rate by the product of transmission probabilities, providing an estimate of how many pairs per second are produced at the source. For quantifying noise performance, we use the CAR, defined as~\cite{halder2008high,krapick2013efficient}
\begin{equation}
\mathrm{CAR} = \frac{C_{\mathrm{true}}}{A_h}
    \label{Eq:CARTrCncd}
\end{equation}

The CAR is a useful figure of merit indicating the signal-to-noise ratio of the heralded photon source, largely independent of detection efficiency. In the ideal case of purely single-pair events, CAR would be very high; multi-pair emission or background counts tend to lower the CAR. We expect that at low pump powers (where multi-pair probability is negligible), CAR will be high, and as the pump power increases, CAR will drop as accidentals (from multiple pairs) grow faster than true pairs.

In a cascaded SHG$\to$SPDC architecture, the SPDC pump at 780~nm is generated by SHG
from the fundamental pump at 1560~nm. In the low-gain regime, the SHG power scales
approximately as $P_{780} \propto P_{1560}^{2}$, and the SPDC pair rate scales as
$R_{\mathrm{pair}} \propto P_{780}$. Therefore, with respect to the fundamental pump,
$R_{\mathrm{pair}} \propto P_{1560}^{2}$. The detected singles scale as
$S'_s, S'_i \propto R_{\mathrm{pair}} \propto P_{1560}^{2}$, while accidentals scale
approximately as $A_h \propto S'_s S'_i \propto P_{1560}^{4}$. Consequently,

\begin{equation}
\mathrm{CAR} \propto C_{\mathrm{true}}/A_h \propto 1/P_{1560}^{2}
    \label{Eq:TrueConic}
\end{equation}

which is consistent with
the trend observed in Fig.~\ref{fig:car_heralding}(a). Since in our cascaded scheme $P_{\omega}^{2}$ itself grows as $P_{\omega}^{2}$ (the fundamental pump power)~\cite{aboussouan2010high,cordier2020raman}, it follows that $\mathrm{CAR} \propto 1/P_{\omega}^{2}$. We will observe this trend in our results.

Finally, to measure the Franson two-photon interference, we vary the relative phase $\phi$ between the signal and idler interferometers and record the coincident count rate as a function of $\phi$. Experimentally, we set one interferometer (e.g., signal) as a reference and scan the temperature of the other (idler) to sweep its phase. The coincidence histograms at each phase setting show three peaks corresponding to the arrival-time differences when the photons take short-short, long-long, or one short/one long paths as presented in Fig.~\ref{fig:g2_fit}(b). Only the central peak (SS/LL) corresponds to indistinguishable two-photon paths and thus shows interference, while the side peaks (SL or LS), where one photon experiences the delay and the other does not, serve as a baseline and remain constant with phase~\cite{franson1989bell,marcikic2004distribution}. We integrate the counts under the central peak (within a $\pm 100$~ps window around zero delay) to obtain the coincidence rate as a function of phase, $C(\phi)$. The raw Franson visibility is then obtained from the maximum and minimum coincidence rates:
\begin{equation}
V_{\mathrm{raw}} = \frac{C_{\max} - C_{\min}}{C_{\max} + C_{\min}}
    \label{Eq:VisComp}
\end{equation}

We also compute a net visibility (background-corrected) by first subtracting accidental counts from each measurement before taking the contrast, and a fitted visibility by fitting the $C(\phi)$ data to a sinusoidal function and extracting the modulation depth. High visibility indicates a high degree of two-photon coherence and entanglement quality; 100\% visibility (the ideal limit) is only approached when the photon pairs are perfectly indistinguishable and noise-free.

\subsubsection{Energy--time entanglement and Franson coincidence model}
Under continuous-wave pumping with pump coherence time $\tau_p$ much longer than the
single-photon coherence time $\tau_c$, the SPDC output exhibits energy--time entanglement.
After propagation through two unbalanced Mach--Zehnder interferometers (uMZIs) with delay
$\Delta t$ satisfying $\tau_c \ll \Delta t \ll \tau_p$, the coincidence histogram versus
arrival-time difference $\tau$ contains three well-separated peaks associated with the
path combinations SL, LS (side peaks at $\tau \approx \pm \Delta t$) and SS/LL (central peak at
$\tau \approx 0$). The SL and LS contributions are distinguishable and therefore phase independent,
whereas the SS and LL amplitudes are indistinguishable within $\tau_p$ and interfere, producing a
phase-dependent modulation of the central peak (Franson interference)~\cite{franson1989bell,tittel2000quantum,halder2008high}.

For fitting and visualization, we model the (optionally accidental-subtracted) coincidence histogram
as the sum of three Gaussian peaks with a phase-dependent central amplitude,
\begin{equation}
\label{eq:franson_hist_model}
\begin{aligned}
H(\tau,\phi)=H_{\rm SL}\exp\!\left[-\frac{(\tau+\Delta t)^2}{2\sigma^2}\right]
&+H_{\rm LS}\exp\!\left[-\frac{(\tau-\Delta t)^2}{2\sigma^2}\right]\\
&+H_{0}\!\left[1+V\cos(\phi+\phi_0)\right]\exp\!\left[-\frac{\tau^{2}}{2\sigma^{2}}\right]
\;(+\,H_{\rm acc})
\end{aligned}
\end{equation}
where $\sigma$ is the effective timing width (set by the convolution of detector timing jitter and
the biphoton wavepacket), $H_{\rm SL}$ and $H_{\rm LS}$ are the side-peak amplitudes, $H_0$ is the
mean central-peak amplitude, and $V$ is the Franson visibility. The optional constant term
$H_{\rm acc}$ captures a flat accidental background when accidentals are not subtracted explicitly.
The phase-dependent coincidence rate $C(\phi)$ is obtained by integrating the
central SS/LL peak of the coincidence histogram $H(\tau,\phi)$ within a fixed
temporal window $|\tau|\leq \tau_{\mathrm{win}}$ around zero delay,
\begin{equation}
C(\phi) = \int_{-\tau_{\mathrm{win}}}^{+\tau_{\mathrm{win}}} H(\tau,\phi)\, d\tau.
\label{phase-dependent coincidence}
\end{equation}

\begin{figure}[t]
\centering

% ---------- Left image ----------
\begin{subfigure}[t]{0.5\textwidth}
    \centering
    \includegraphics[width=\linewidth, height=6.0cm, keepaspectratio]{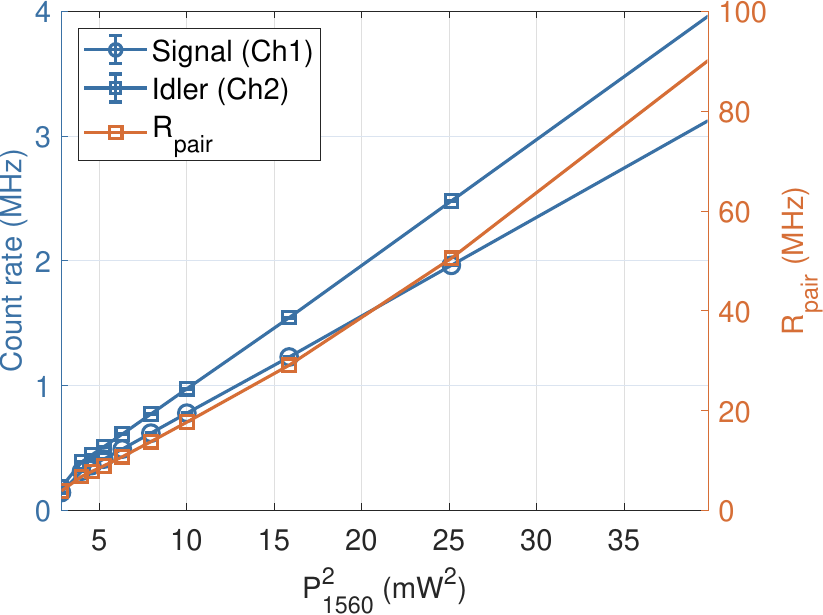}
    \caption{} % This automatically adds "(a)"
    \label{fig:pair_rate_vs_power}
\end{subfigure}
\hfill
% ---------- Right image ----------
\begin{subfigure}[t]{0.47\textwidth}
    \centering
    \includegraphics[width=\linewidth, height=6cm, keepaspectratio]{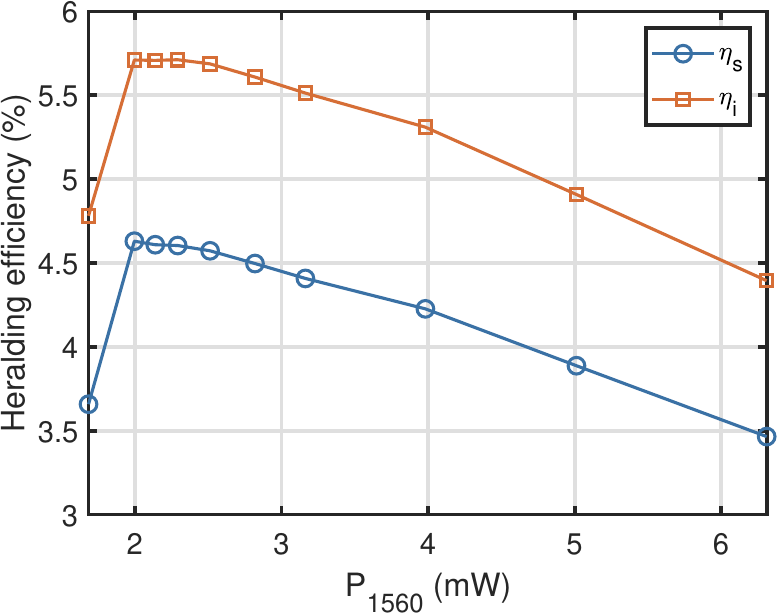}
    \caption{} % This automatically adds "(b)"
    \label{fig:HerladingEff}
\end{subfigure}

\caption{(a)
Signal ($S'_s$) and idler ($S'_i$) singles rates together with the inferred 
pair-generation rate $R_{\mathrm{pair}}$ as a function of the squared pump 
power $P_{1560}^{2}$. The linear dependence confirms the expected quadratic 
scaling of the SPDC pair rate with the fundamental pump power in the cascaded 
SHG$\rightarrow$SPDC architecture. (b) Heralding efficiencies $\eta_s$ and $\eta_i$ as a function of pump power 
$P_{1560}$. The efficiencies peak near 2~mW and decrease with increasing pump 
power due to rising multi-pair emission probability and mild detector 
saturation. The close agreement between $\eta_s$ and $\eta_i$ confirms balanced 
optical throughput in the signal and idler arms.}

\label{fig:CR_RR}
\end{figure}

\section{Results and Discussion}

\subsection{Source Brightness, Heralding Efficiency, and Noise Performance}

Using the definitions and corrections described in Sec.~\ref{Meas&Anlss} [Eqs.~(\ref{Eq:SingletRates})--(\ref{Eq:CARTrCncd})], we evaluate the corrected singles rates $S'_s$ and $S'_i$, the inferred internal pair generation rate $R_{\mathrm{pair}}$, the heralding efficiencies $\eta_s$ and $\eta_i$, and the CAR. Figure~\ref{fig:CR_RR}(a) shows $S'_s$, $S'_i$, and $R_{\mathrm{pair}}$ as functions of $P_{1560}^{2}$. All three quantities scale linearly with $P_{1560}^{2}$ over the measured range, consistent with the expected quadratic dependence of the pair production rate on the fundamental pump power in a cascaded SHG$\rightarrow$SPDC architecture. The linear fits exhibit no evidence of saturation or additional parasitic nonlinearities at the explored pump levels.

Figure~\ref{fig:CR_RR}(b) shows the heralding efficiencies $\eta_s$ and $\eta_i$ [Eqs.~(\ref{Eq:HeraldingEff_Signal})--(\ref{Eq:HeraldingEff_Idler})] versus pump power. We obtain $\eta_s \approx \eta_i$ in the range 3.5--4.8\%, with a maximum of $\sim$4.8\% near $P_{1560}\approx 2$~mW and a gradual decrease at higher powers. The agreement between $\eta_s$ and $\eta_i$ within experimental uncertainty provides an internal consistency check for balanced channel losses and the calibration of the detection and correction procedures~\cite{klyshko1980use,aboussouan2010high}. While these system heralding efficiencies are below the highest values reported for fully optimized or monolithically integrated sources, they are representative of fiber-coupled waveguide SPDC systems employing external DWDM
filtering and standalone fiber-coupled detection~\cite{halder2008high,takesue2005generation}. In the present setup, the dominant limitations are fiber coupling losses at the PPLN module, the $\sim$5.4~dB insertion loss of the DWDM filters, and the detector quantum efficiency ($\sim$80\%). These figures are therefore best interpreted as end-to-end system efficiencies rather than intrinsic source limitations, and they could be improved through reduced coupling loss and/or integrated wavelength demultiplexing~\cite{aboussouan2010high,krapick2013efficient}.

At the lowest pump powers ($\lesssim 0.5$~mW), $\eta$ decreases slightly as the detected pair rate approaches the noise floor, increasing the relative weight of spurious heralds~\cite{halder2008high,takesue2005generation}. At higher powers, the gradual reduction in $\eta$ is consistent with an increasing contribution of multi-pair emission and mild detector saturation effects~\cite{krapick2013efficient,davanco2012telecommunications}.

\begin{figure}[t]
\centering
\begin{subfigure}[t]{0.40\linewidth}
  \centering
  \includegraphics[width=\linewidth,height=5.6cm,keepaspectratio]{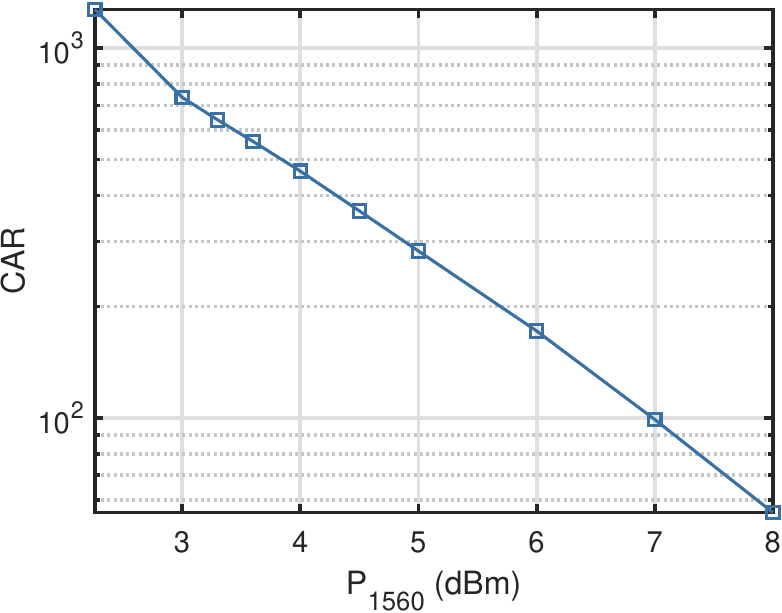}
  \caption{}
\end{subfigure}\hfill
\begin{subfigure}[t]{0.50\linewidth}
  \centering
  \includegraphics[width=\linewidth,height=6.0cm,keepaspectratio]{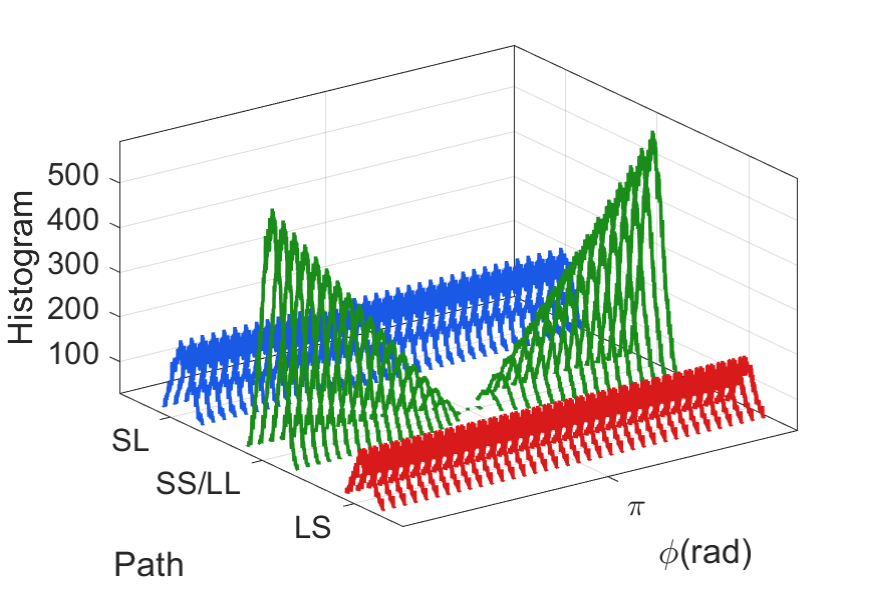}
  \caption{}
\end{subfigure}
\caption{(a) Coincidence-to-accidental ratio (CAR) as a function of the
    fundamental pump power \(P_{1560}\). The data are plotted on a logarithmic
    scale and follow the expected \(\mathrm{CAR}\propto 1/P_\omega^2\)
    dependence (solid line) derived for the cascaded SHG–SPDC process,
    corresponding to a slope of approximately \(-0.2\) decades per dB.
    The CAR decreases from \(\sim 10^{3}\) at the lowest pump powers to
    \(\sim 10^{2}\) at \(P_{1560}=8~\mathrm{dBm}\), indicating the
    increasing impact of multi-pair events at higher brightness.
(b) Coincidence histograms measured for
constructive (\(\phi=0\)) and destructive (\(\phi=\pi\)) two-photon
interference. The central peak (SS/LL) exhibits the expected
phase-dependent modulation, increasing for \(\phi=0\) and decreasing for
\(\phi=\pi\), while the side peaks (SL and LS) remain constant under
phase variation, as they correspond to distinguishable path
combinations. This invariance of the SL/LS peaks provides a direct
baseline for identifying genuine energy--time interference in the
central SS/LL contribution.}
\label{fig:car_heralding}
\end{figure}

Noise performance is quantified by the CAR [Eq.~(\ref{Eq:CARTrCncd})], shown in Fig.~\ref{fig:car_heralding}(a). At $P_{1560}\approx 1$~mW (0~dBm), we measure $\mathrm{CAR}\sim 10^{3}$, decreasing to $\sim 10^{2}$ at the highest pump powers ($\sim$6--7~mW, 8~dBm). Across most of the range, the data follow the scaling predicted by the low-gain analysis in Sec.~2.2 (CAR $\propto P^{-2}$), indicating that accidentals are dominated by multi-pair emission rather than dark counts or extrinsic background~\cite{aboussouan2010high,cordier2020raman,krapick2013efficient,davanco2012telecommunications}. A slight deviation from the ideal $1/P^{2}$ trend appears only at the lowest powers, consistent with a background-limited floor when true coincidences become comparable to fixed accidental contributions~\cite{takesue2005generation,harder2013optimized}.

Overall, maintaining CAR $>100$ at multi-milliwatt pump powers indicates low-noise operation at useful brightness. For context, comparable CAR values have been reported in telecom-band waveguide sources under narrow filtering and/or gated detection, whereas other platforms (e.g., silicon nanophotonics) are often limited to substantially lower CAR at practical pair rates due to Raman noise and related nonlinear processes~\cite{krapick2013efficient,cordier2020raman,davanco2012telecommunications,spring2013chip}. In the present regime, CAR in the hundreds or above implies that accidental coincidences contribute only weakly to visibility reduction, consistent with the high-contrast interference reported below~\cite{halder2008high,krapick2013efficient}.

\subsection{Two-Photon Franson Interference Visibility}

Two-photon interference is analyzed using the Franson geometry and coincidence-histogram model summarized in Sec.~\ref{Meas&Anlss} (Energy--time entanglement and Franson coincidence model). Figure~\ref{fig:car_heralding}(b) shows representative coincidence histograms for two extreme phase settings, corresponding to constructive ($\phi\approx 0$) and destructive ($\phi\approx \pi$) interference. Consistent with the Franson regime ($\tau_c \ll \Delta t \ll \tau_p$), the side peaks (SL and LS) are phase independent (though not necessarily equal in amplitude due to unequal channel transmission), whereas the central SS/LL peak exhibits strong phase-dependent modulation~\cite{franson1989bell,marcikic2004distribution}. The suppression of the central peak near $\phi\approx \pi$ provides direct evidence of high-contrast destructive two-photon interference and hence high energy--time indistinguishability~\cite{marcikic2004distribution,tittel2000quantum}.

\begin{figure}[t]
\centering
\begin{subfigure}[t]{0.5\linewidth}
  \centering
  \includegraphics[width=\linewidth,height=6cm,keepaspectratio]{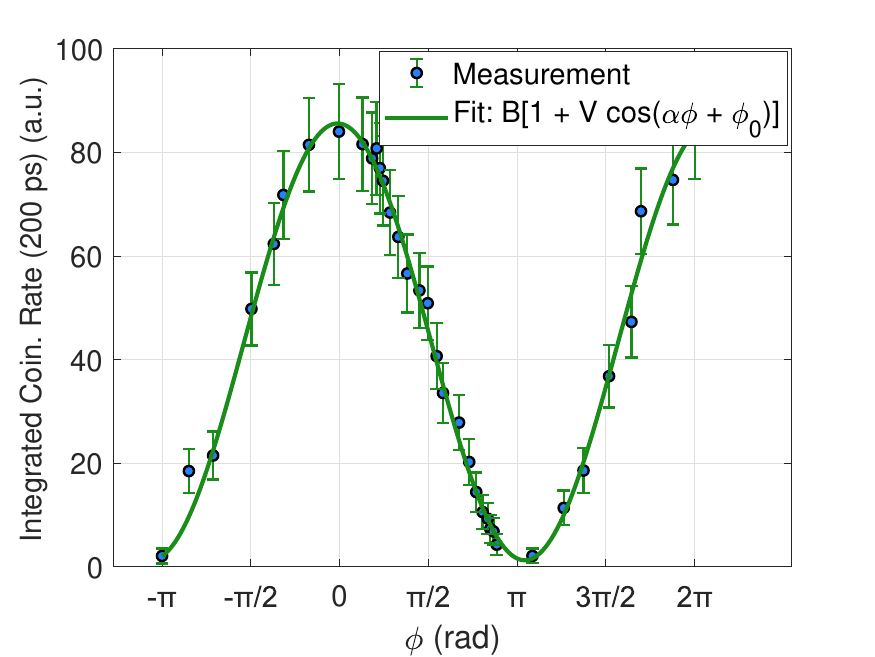}
  \caption{}
\end{subfigure}\hfill
\begin{subfigure}[t]{0.45\linewidth}
  \centering
  \includegraphics[width=\linewidth,height=5.6cm,keepaspectratio]{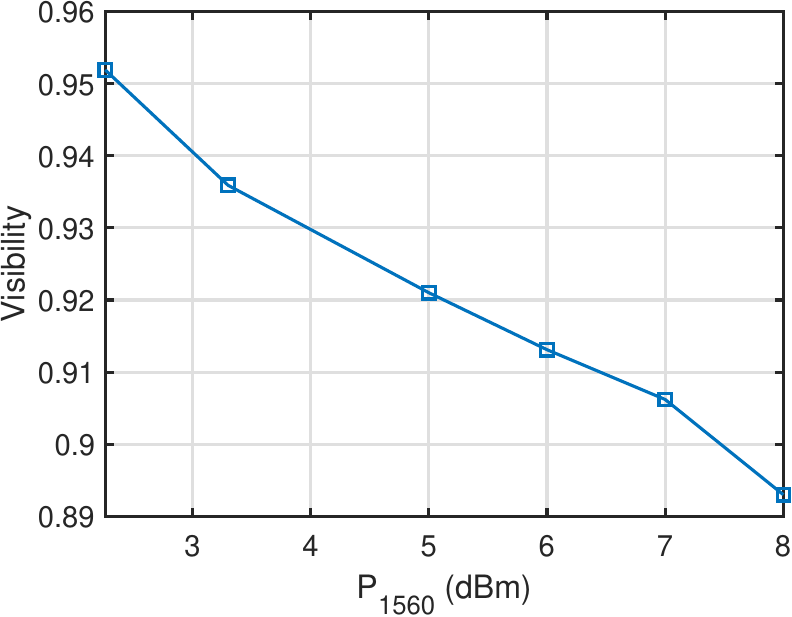}
  \caption{}
\end{subfigure}
\caption{(a) Central-peak coincidence rate versus phase; 
the sinusoidal fit $C(\phi)=B[1+V\cos(\alpha\phi+\phi_0)]$ 
yields $V=97.1\%$. 
Raw visibility from (a): $95.2\%$; background-corrected: $95.6\%$. (b) Raw two-photon interference visibility as a function of the fundamental pump power $P_{1560}$. As the pump power increases, the visibility decreases from
about \(95\%\) to \(89\%\), consistent with the growing contribution of
multi-pair emission and accidental coincidences at higher source
brightness.}
\label{fig:franson}
\end{figure}

We quantify the interference fringe by integrating the SS/LL peak within the fixed window defined in Sec.~\ref{Meas&Anlss} to obtain $C(\phi)$. Figure~\ref{fig:franson}(a) shows $C(\phi)$ for $P_{1560}=1.7$~mW, where the fit visibility is maximal. A sinusoidal fit yields $V=97.1\%$, while the raw and background-corrected visibilities are 95.2\% and 95.6\%, respectively. The small difference between raw and corrected values indicates that accidentals contribute only weakly to fringe washout, consistent with the measured CAR~\cite{halder2008high,krapick2013efficient}. The difference between the fit-derived visibility and the point-estimate (raw/net) visibilities is
consistent with finite sampling and measurement noise; residual deviations from a perfect
sinusoid may arise from slow thermal drift during the scan or small interferometer imbalance.

Figure~\ref{fig:franson}(b) shows the pump-power dependence of the raw visibility. The visibility decreases from $\sim$95\% at $\sim$1~mW to $\sim$89\% at $\sim$6~mW, consistent with increasing multi-pair emission and the resulting phase-independent background contribution at higher brightness~\cite{krapick2013efficient,davanco2012telecommunications,aboussouan2010high}. Notably, visibility remains close to 90\% even at the highest pump powers tested, indicating robust entanglement quality at high pair rates and supporting operation in a regime that balances brightness and interference contrast for network-relevant scenarios.

\begin{table}[t]
\caption{Comparison of representative high-visibility Franson interference demonstrations.}
\centering
\begin{tabular}{p{2.3cm} p{4.0cm} p{5.2cm} p{2.2cm}}
\toprule
Ref. & Platform / Architecture & Analyzer (interferometer; filter / $\Delta t$) & Visibility \\
\midrule

Halder et al.\ [2008] 
& Single PPLN waveguide SPDC 
& Fiber Franson; $\sim$100\,GHz; $\Delta t \sim 1$ ns 
& $\sim$91\% (net) \\

Da Lio et al.\ [2021] 
& MCF transmission + integrated PIC analyzers 
& PIC uMZIs; $\sim$100\,GHz; $\Delta t \sim 0.6$ ns 
& $\sim$94$-$95\% (raw) \\

Zhong et al.\ [2012] 
& PPKTP waveguide SPDC 
& Fiber Franson; $\sim$50\,GHz; $\Delta t \sim 1$ ns 
& $98.2\%\pm0.3\%$ (raw) \\

Zhang et al.\ [2021] 
& Cascaded SHG$\rightarrow$SPDC PPLN 
& Fiber Franson; $\sim$100\,GHz; $\Delta t \sim 0.8$ ns 
& $95.74\%\pm0.86\%$ (fit) \\

Park et al.\ [2019] 
& Warm atomic ensemble 
& Free-space Franson; $<1$ GHz; $\Delta t \sim 1$ ns 
& $99.1\%\pm1.3\%$ (net) \\

Troisi et al.\ [2026] 
& Fiber-based nonlinear Sagnac (cascaded PPLN) 
& Fiber Franson (Michelson, Faraday mirrors); 100\,GHz DWDM; $\Delta t \sim 1$ ns 
& $99\%\pm1\%$ (raw) \\

\textbf{This work} 
& Cascaded PPLN + passive PIC uMZIs 
& PIC uMZIs; 100\,GHz DWDM; $\Delta t \sim 0.8$ ns 
& $95.2\%$ (raw); $95.6\%$ (net); $97.1\%$ (fit) \\

\bottomrule
\end{tabular}
\label{tab:franson_comparison}
\end{table}

To place these results in context (Table~\ref{tab:franson_comparison}), Halder \emph{et al.} reported $\sim$91\% net
visibility using a single PPLN waveguide source with fiber interferometers~\cite{halder2008high}, while
Da~Lio \emph{et al.} achieved 94--95\% raw visibility using multicore-fiber distribution with on-chip
interferometric analysis~\cite{da2021path}. Zhong \emph{et al.} reported a benchmark 98.2\% raw visibility
using a PPKTP waveguide source with stringent narrowband filtering~\cite{zhong2012efficient}, and a
cascaded PPLN approach similar to ours achieved a fitted visibility of $\sim$95.7\% with fiber-based
Franson interferometers~\cite{zhang2021high}. Very recently, Troisi \emph{et al.} demonstrated a fully
fibered nonlinear Sagnac source based on cascaded PPLN waveguides, combining 100~GHz DWDM channelization
with a passively stabilized fiber Franson analyzer (Michelson configuration with Faraday mirrors),
and reported a raw Franson visibility of $99\pm1\%$ on a selected ITU channel pair (19,23)~\cite{Troisi2026Sagnac}.
In a different physical platform, Park \emph{et al.} demonstrated $\sim$99\% net visibility using a warm
Rb ensemble under narrowband conditions~\cite{park2019high}. Our measured visibilities (95.2\% raw,
95.6\% net, and 97.1\% fitted at 1550~nm) therefore place this system among the highest-visibility
photonic-integrated Franson analyzers reported to date in the telecom band, while uniquely combining a
cascaded $\chi^{(2)}$ source architecture with fully passive, thermally tuned PIC interferometers and
DWDM-compatible fiber integration~\cite{zhang2021high,guarda2024quantum}.

\section{Conclusion}

We have realized a compact, fiber-pigtailed energy-time entanglement platform that combines a cascaded PPLN waveguide SPDC source with fully passive photonic integrated interferometers operating in the telecom C-band. The system achieves two-photon Franson interference visibilities above 95\% (95.2\% raw, 95.6\% net, and 97.1\% from fringe fitting) together with a CAR exceeding $10^{3}$ at milliwatt-level pump powers, demonstrating low noise, high indistinguishability, and long-term phase stability without active locking. By leveraging cascaded $\chi^{(2)}$ nonlinear generation for spectral purity and monolithic PIC interferometers for robust phase control, this approach attains state-of-the-art visibility with reduced experimental complexity and full compatibility with DWDM-based fiber networks. Future improvements in spectral filtering, on-chip demultiplexing, and non-post-selected interferometric schemes could further enhance heralding efficiency and extend applicability to device-independent protocols. Overall, these results establish cascaded PPLN sources combined with passive integrated interferometers as a practical and scalable route toward high-visibility, field-deployable quantum entanglement distribution.

%\bibliographystyle{unsrtnat}
%\bibliography{references.bib}

%\bibliographystyle{unsrtnat}
%\bibliography{references}  %%% Uncomment this line and comment out the ``thebibliography'' section below to use the external .bib file (using bibtex) .


\begin{thebibliography}{99}

\bibitem[Paudel et~al.(2023)Paudel, Crawford, Lee, Shugayev, Leuenberger, Syamlal, Ohodnicki, Lu, Mollot, and Duan]{paudel2023quantum}
Hari P.~Paudel, Scott E.~Crawford, Yueh-Lin Lee, Roman A.~Shugayev, Michael N.~Leuenberger, Madhava Syamlal, Paul R.~Ohodnicki, Ping Lu, Darren Mollot, and Yuhua Duan.
Quantum communication networks for energy applications: Review and perspective.
\emph{Advanced Quantum Technologies}, 6(10):2300096, 2023.

\bibitem[Kimble(2008)]{kimble2008quantum}
H.~Jeff Kimble.
The quantum internet.
\emph{Nature}, 453(7198):1023--1030, 2008.

\bibitem[Tagliavacche et~al.(2025)Tagliavacche, Borghi, Guarda, Ribezzo, Liscidini, Bacco, Galli, and Bajoni]{tagliavacche2025frequency}
Noemi Tagliavacche, Massimo Borghi, Giulia Guarda, Domenico Ribezzo, Marco Liscidini, Davide Bacco, Matteo Galli, and Daniele Bajoni.
Frequency-bin entanglement-based quantum key distribution.
\emph{npj Quantum Information}, 11(1):60, 2025.

\bibitem[Laucht et~al.(2021)Laucht, Hohls, Ubbelohde, Gonzalez-Zalba, Reilly, Stobbe, Schr{\"o}der, Scarlino, Koski, Dzurak, and others]{laucht2021roadmap}
Arne Laucht, Frank Hohls, Niels Ubbelohde, M.~Fernando Gonzalez-Zalba, David J.~Reilly, S{\o}ren Stobbe, Tim Schr{\"o}der, Pasquale Scarlino, Jonne V.~Koski, Andrew Dzurak, and others.
Roadmap on quantum nanotechnologies.
\emph{Nanotechnology}, 32(16):162003, 2021.

\bibitem[Politi et~al.(2009)Politi, Matthews, Thompson, and O'Brien]{politi2009integrated}
Alberto Politi, Jonathan C.~F. Matthews, Mark G.~Thompson, and Jeremy L.~O'Brien.
Integrated quantum photonics.
\emph{IEEE Journal of Selected Topics in Quantum Electronics}, 15(6):1673--1684, 2009.

\bibitem[Silverstone et~al.(2016)Silverstone, Bonneau, O'Brien, and Thompson]{silverstone2016silicon}
Joshua W.~Silverstone, Damien Bonneau, Jeremy L.~O'Brien, and Mark G.~Thompson.
Silicon quantum photonics.
\emph{IEEE Journal of Selected Topics in Quantum Electronics}, 22(6):390--402, 2016.

\bibitem[Gisin and Thew(2007)Gisin and Thew]{gisin2007quantum}
Nicolas Gisin and Rob Thew.
Quantum communication.
\emph{Nature Photonics}, 1(3):165--171, 2007.

\bibitem[Ribordy et~al.(2000)Ribordy, Brendel, Gautier, Gisin, and Zbinden]{ribordy2000long}
Gr{\'e}goire Ribordy, J{\"u}rgen Brendel, Jean-Daniel Gautier, Nicolas Gisin, and Hugo Zbinden.
Long-distance entanglement-based quantum key distribution.
\emph{Physical Review A}, 63(1):012309, 2000.

\bibitem[Xavier et~al.(2025)Xavier, Larsson, Villoresi, Vallone, and Cabello]{xavier2025energy}
Guilherme B.~Xavier, Jan-{\AA}ke Larsson, Paolo Villoresi, Giuseppe Vallone, and Ad{\'a}n Cabello.
Energy-time and time-bin entanglement: past, present and future.
\emph{npj Quantum Information}, 11(1):129, 2025.

\bibitem[Olislager et~al.(2010)Olislager, Cussey, Nguyen, Emplit, Massar, Merolla, and Huy]{olislager2010frequency}
Laurent Olislager, Johann Cussey, Anh Tuan Nguyen, Philippe Emplit, Serge Massar, J.-M. Merolla, and K.~Phan Huy.
Frequency-bin entangled photons.
\emph{Physical Review A}, 82(1):013804, 2010.

\bibitem[Lukens and Lougovski(2016)Lukens and Lougovski]{lukens2016frequency}
Joseph M.~Lukens and Pavel Lougovski.
Frequency-encoded photonic qubits for scalable quantum information processing.
\emph{Optica}, 4(1):8--16, 2016.

\bibitem[Kues et~al.(2017)Kues, Reimer, Wetzel, Roztocki, Little, Chu, Hansson, Viktorov, Moss, and Morandotti]{kues2017passively}
Michael Kues, Christian Reimer, Benjamin Wetzel, Piotr Roztocki, Brent E.~Little, Sai T.~Chu, Tobias Hansson, Evgeny A.~Viktorov, David J.~Moss, and Roberto Morandotti.
Passively mode-locked laser with an ultra-narrow spectral width.
\emph{Nature Photonics}, 11(3):159--162, 2017.

\bibitem[Da Lio et~al.(2019)Da Lio, Bacco, Cozzolino, Biagi, Arge, Larsen, Rottwitt, Ding, Zavatta, and Oxenl{\o}we]{da2019stable}
Beatrice Da Lio, Davide Bacco, Daniele Cozzolino, Nicola Biagi, Tummas Napoleon Arge, Emil Larsen, Karsten Rottwitt, Yunhong Ding, Alessandro Zavatta, and Leif Katsuo Oxenl{\o}we.
Stable transmission of high-dimensional quantum states over a 2-km multicore fiber.
\emph{IEEE Journal of Selected Topics in Quantum Electronics}, 26(4):1--8, 2019.

\bibitem[Wehner et~al.(2018)Wehner, Elkouss, and Hanson]{wehner2018quantum}
Stephanie Wehner, David Elkouss, and Ronald Hanson.
Quantum internet: A vision for the road ahead.
\emph{Science}, 362(6412):eaam9288, 2018.

\bibitem[Wang et~al.(2020)Wang, Sciarrino, Laing, and Thompson]{wang2020integrated}
Jianwei Wang, Fabio Sciarrino, Anthony Laing, and Mark G.~Thompson.
Integrated photonic quantum technologies.
\emph{Nature Photonics}, 14(5):273--284, 2020.

\bibitem[Franson(1989)]{franson1989bell}
James D.~Franson.
Bell inequality for position and time.
\emph{Physical Review Letters}, 62(19):2205, 1989.

\bibitem[Tittel et~al.(2000)Tittel, Brendel, Zbinden, and Gisin]{tittel2000quantum}
Wolfgang Tittel, J{\"u}rgen Brendel, Hugo Zbinden, and Nicolas Gisin.
Quantum cryptography using entangled photons in energy-time Bell states.
\emph{Physical Review Letters}, 84(20):4737, 2000.

\bibitem[Marcikic et~al.(2004)Marcikic, De Riedmatten, Tittel, Zbinden, Legr{\'e}, and Gisin]{marcikic2004distribution}
Ivan Marcikic, Hugues De Riedmatten, Wolfgang Tittel, Hugo Zbinden, Matthieu Legr{\'e}, and Nicolas Gisin.
Distribution of time-bin entangled qubits over 50 km of optical fiber.
\emph{Physical Review Letters}, 93(18):180502, 2004.

\bibitem[Fan et~al.(2023)Fan, Luo, Zhang, Li, Liu, Wang, Zhang, Deng, Wang, Song, and others]{fan2023energy}
Yun-Ru Fan, Yue Luo, Zi-Chang Zhang, Yun-Bo Li, Sheng Liu, Dong Wang, De-Chao Zhang, Guang-Wei Deng, You Wang, Hai-Zhi Song, and others.
Energy-time entanglement coexisting with fiber-optical communication in the telecom C band.
\emph{Physical Review A}, 108(2):L020601, 2023.

\bibitem[Kwiat et~al.(1993)Kwiat, Steinberg, and Chiao]{kwiat1993high}
Paul G.~Kwiat, Aephraim M.~Steinberg, and Raymond Y.~Chiao.
High-visibility interference in a Bell-inequality experiment for energy and time.
\emph{Physical Review A}, 47(4):R2472, 1993.

\bibitem[Rarity and Tapster(1990)Rarity and Tapster]{rarity1990experimental}
J.~G. Rarity and P.~R. Tapster.
Experimental violation of Bell's inequality based on phase and momentum.
\emph{Physical Review Letters}, 64(21):2495, 1990.

\bibitem[Brendel et~al.(1991)Brendel, Mohler, and Martienssen]{brendel1991time}
J.~Brendel, E.~Mohler, and W.~Martienssen.
Time-resolved dual-beam two-photon interferences with high visibility.
\emph{Physical Review Letters}, 66(9):1142, 1991.

\bibitem[Viciani et~al.(2004)Viciani, Zavatta, and Bellini]{viciani2004nonlocal}
Silvia Viciani, Alessandro Zavatta, and Marco Bellini.
Nonlocal modulations on the temporal and spectral profiles of an entangled photon pair.
\emph{Physical Review A}, 69(5):053801, 2004.

\bibitem[Zavatta et~al.(2006)Zavatta, D'Angelo, Parigi, and Bellini]{zavatta2006remote}
Alessandro Zavatta, Milena D'Angelo, Valentina Parigi, and Marco Bellini.
Remote preparation of arbitrary time-encoded single-photon ebits.
\emph{Physical Review Letters}, 96(2):020502, 2006.

\bibitem[Cozzolino et~al.(2019)Cozzolino, Da Lio, Bacco, and Oxenl{\o}we]{cozzolino2019high}
Daniele Cozzolino, Beatrice Da Lio, Davide Bacco, and Leif Katsuo Oxenl{\o}we.
High-dimensional quantum communication: benefits, progress, and future challenges.
\emph{Advanced Quantum Technologies}, 2(12):1900038, 2019.

\bibitem[Vagniluca et~al.(2020)Vagniluca, Da Lio, Rusca, Cozzolino, Ding, Zbinden, Zavatta, Oxenl{\o}we, and Bacco]{vagniluca2020efficient}
Ilaria Vagniluca, Beatrice Da Lio, Davide Rusca, Daniele Cozzolino, Yunhong Ding, Hugo Zbinden, Alessandro Zavatta, Leif K.~Oxenl{\o}we, and Davide Bacco.
Efficient time-bin encoding for practical high-dimensional quantum key distribution.
\emph{Physical Review Applied}, 14(1):014051, 2020.

\bibitem[Da Lio et~al.(2021)Da Lio, Cozzolino, Biagi, Ding, Rottwitt, Zavatta, Bacco, and Oxenl{\o}we]{da2021path}
Beatrice Da Lio, Daniele Cozzolino, Nicola Biagi, Yunhong Ding, Karsten Rottwitt, Alessandro Zavatta, Davide Bacco, and Leif K.~Oxenl{\o}we.
Path-encoded high-dimensional quantum communication over a 2-km multicore fiber.
\emph{npj Quantum Information}, 7(1):63, 2021.

\bibitem[Bacco et~al.(2019)Bacco, Da Lio, Cozzolino, Da Ros, Guo, Ding, Sasaki, Aikawa, Miki, Terai, and others]{bacco2019boosting}
Davide Bacco, Beatrice Da Lio, Daniele Cozzolino, Francesco Da Ros, Xueshi Guo, Yunhong Ding, Yusuke Sasaki, Kazuhiko Aikawa, Shigehito Miki, Hirotaka Terai, and others.
Boosting the secret key rate in a shared quantum and classical fibre communication system.
\emph{Communications Physics}, 2(1):140, 2019.

\bibitem[Halder et~al.(2008)Halder, Beveratos, Thew, Jorel, Zbinden, and Gisin]{halder2008high}
Matth{\"a}us Halder, Alexios Beveratos, Robert T.~Thew, Corentin Jorel, Hugo Zbinden, and Nicolas Gisin.
High coherence photon pair source for quantum communication.
\emph{New Journal of Physics}, 10(2):023027, 2008.

\bibitem[Zhang et~al.(2021)Zhang, Yuan, Shen, Yu, Zhang, Wang, Li, Wang, Deng, Wang, and others]{zhang2021high}
Zichang Zhang, Chenzhi Yuan, Si Shen, Hao Yu, Ruiming Zhang, Heqing Wang, Hao Li, You Wang, Guangwei Deng, Zhiming Wang, and others.
High-performance quantum entanglement generation via cascaded second-order nonlinear processes.
\emph{npj Quantum Information}, 7(1):123, 2021.

\bibitem[Park et~al.(2019)Park, Kim, Kim, and Moon]{park2019high}
Jiho Park, Danbi Kim, Heonoh Kim, and Han Seb Moon.
High-visibility Franson interference of time--energy entangled photon pairs from warm atomic ensemble.
\emph{Optics Letters}, 44(15):3681--3684, 2019.

\bibitem[Boyd et~al.(2008)Boyd, Gaeta, and Giese]{boyd2008nonlinear}
Robert W.~Boyd, Alexander L.~Gaeta, and Enno Giese.
Nonlinear optics.
In \emph{Springer Handbook of Atomic, Molecular, and Optical Physics}, pp.~1097--1110.
Springer, 2008.

\bibitem[Fejer et~al.(2002)Fejer, Magel, Jundt, and Byer]{fejer2002quasi}
Martin M.~Fejer, G.~A. Magel, Dieter H.~Jundt, and Robert L.~Byer.
Quasi-phase-matched second harmonic generation: tuning and tolerances.
\emph{IEEE Journal of Quantum Electronics}, 28(11):2631--2654, 2002.

\bibitem[Myers et~al.(1995)Myers, Eckardt, Fejer, Byer, Bosenberg, and Pierce]{myers1995quasi}
Lawrence Edward Myers, R.~C. Eckardt, M.~M. Fejer, R.~L. Byer, W.~R. Bosenberg, and J.~W. Pierce.
Quasi-phase-matched optical parametric oscillators in bulk periodically poled LiNbO$_3$.
\emph{Journal of the Optical Society of America B}, 12(11):2102--2116, 1995.

\bibitem[Guarda et~al.(2024)Guarda, Ribezzo, Occhipinti, Zavatta, and Bacco]{guarda2024quantum}
Giulia Guarda, Domenico Ribezzo, Tommaso Occhipinti, Alessandro Zavatta, and Davide Bacco.
Quantum key distribution with a borosilicate-glass-matrix integrated photonic receiver.
\emph{Physical Review A}, 110(4):042605, 2024.

\bibitem[Sibson et~al.(2017)Sibson, Erven, Godfrey, Miki, Yamashita, Fujiwara, Sasaki, Terai, Tanner, Natarajan, and others]{sibson2017chip}
Philip Sibson, Chris Erven, Mark Godfrey, Shigehito Miki, Taro Yamashita, Mikio Fujiwara, Masahide Sasaki, Hirotaka Terai, Michael G.~Tanner, Chandra M.~Natarajan, and others.
Chip-based quantum key distribution.
\emph{Nature Communications}, 8(1):13984, 2017.

\bibitem[Jin et~al.(2019)Jin, Bourgoin, Tannous, Agne, Pugh, Kuntz, Higgins, and Jennewein]{jin2019genuine}
Jeongwan Jin, Jean-Philippe Bourgoin, Ramy Tannous, Sascha Agne, Christopher J.~Pugh, Katanya B.~Kuntz, Brendon L.~Higgins, and Thomas Jennewein.
Genuine time-bin-encoded quantum key distribution over a turbulent depolarizing free-space channel.
\emph{Optics Express}, 27(26):37214--37223, 2019.

\bibitem[Hong and Mandel(1986)Hong and Mandel]{hong1986experimental}
C.~K. Hong and Leonard Mandel.
Experimental realization of a localized one-photon state.
\emph{Physical Review Letters}, 56(1):58, 1986.

\bibitem[Klyshko(1980)]{klyshko1980use}
D.~N. Klyshko.
Use of two-photon light for absolute calibration of photoelectric detectors.
\emph{Soviet Journal of Quantum Electronics}, 10(9):1112, 1980.

\bibitem[Aboussouan et~al.(2010)Aboussouan, Alibart, Ostrowsky, Baldi, and Tanzilli]{aboussouan2010high}
Pierre Aboussouan, Olivier Alibart, Daniel B.~Ostrowsky, Pascal Baldi, and S{\'e}bastien Tanzilli.
High-visibility two-photon interference at a telecom wavelength using picosecond-regime separated sources.
\emph{Physical Review A}, 81(2):021801, 2010.

\bibitem[Mandel and Wolf(1995)Mandel and Wolf]{mandel1995optical}
Leonard Mandel and Emil Wolf.
\emph{Optical Coherence and Quantum Optics}.
Cambridge University Press, 1995.

\bibitem[Goodman(2015)]{goodman2015statistical}
Joseph W.~Goodman.
\emph{Statistical Optics}.
John Wiley \& Sons, 2015.

\bibitem[Knoll(2010)]{knoll2010radiation}
Glenn F.~Knoll.
\emph{Radiation Detection and Measurement}.
John Wiley \& Sons, 2010.

\bibitem[Shi et~al.(2024)Shi, Mohanraj, Dhyani, Baiju, Wang, Sun, Zhou, Paterova, Leong, and Zhu]{shi2024efficient}
Xiaodong Shi, Sakthi Sanjeev Mohanraj, Veerendra Dhyani, Angela Anna Baiju, Sihao Wang, Jiapeng Sun, Lin Zhou, Anna Paterova, Victor Leong, and Di Zhu.
Efficient photon-pair generation in layer-poled lithium niobate nanophotonic waveguides.
\emph{Light: Science \& Applications}, 13(1):282, 2024.

\bibitem[Hellwig(2011)]{hellwig2011nonlinear}
Ansgar Hellwig.
\emph{Nonlinear Optical and Photorefractive Properties of Periodically Poled Channel Waveguides in Lithium Niobate}.
PhD thesis, Universit{\"a}tsbibliothek, 2011.

\bibitem[Broquin and Honkanen(2021)Broquin and Honkanen]{broquin2021integrated}
Jean-Emmanuel Broquin and Seppo Honkanen.
Integrated photonics on glass: A review of the ion-exchange technology achievements.
\emph{Applied Sciences}, 11(10):4472, 2021.

\bibitem[Spring et~al.(2013)Spring, Salter, Metcalf, Humphreys, Moore, Thomas-Peter, Barbieri, Jin, Langford, Kolthammer, and others]{spring2013chip}
Justin B.~Spring, Patrick S.~Salter, Benjamin J.~Metcalf, Peter C.~Humphreys, Merritt Moore, Nicholas Thomas-Peter, Marco Barbieri, Xian-Min Jin, Nathan K.~Langford, W.~Steven Kolthammer, and others.
On-chip low loss heralded source of pure single photons.
\emph{Optics Express}, 21(11):13522--13532, 2013.

\bibitem[Krapick et~al.(2013)Krapick, Herrmann, Quiring, Brecht, Suche, and Silberhorn]{krapick2013efficient}
Stephan Krapick, Harald Herrmann, Viktor Quiring, Benjamin Brecht, Hubertus Suche, and Christine Silberhorn.
An efficient integrated two-color source for heralded single photons.
\emph{New Journal of Physics}, 15(3):033010, 2013.

\bibitem[Tanzilli et~al.(2001)Tanzilli, De Riedmatten, Tittel, Zbinden, Baldi, De Micheli, Ostrowsky, and Gisin]{tanzilli2001highly}
S{\'e}bastien Tanzilli, Hugues De Riedmatten, Wolfgang Tittel, Hugo Zbinden, Pascal Baldi, Marc De Micheli, Daniel Barry Ostrowsky, and Nicolas Gisin.
Highly efficient photon-pair source using periodically poled lithium niobate waveguide.
\emph{Electronics Letters}, 37(1):26--28, 2001.

\bibitem[Mosley et~al.(2008)Mosley, Lundeen, Smith, Wasylczyk, U'Ren, Silberhorn, and Walmsley]{mosley2008heralded}
Peter J.~Mosley, Jeff S.~Lundeen, Brian J.~Smith, Piotr Wasylczyk, Alfred B.~U'Ren, Christine Silberhorn, and Ian A.~Walmsley.
Heralded generation of ultrafast single photons in pure quantum states.
\emph{Physical Review Letters}, 100(13):133601, 2008.

\bibitem[Takesue and Inoue(2005)Takesue and Inoue]{takesue2005generation}
Hiroki Takesue and Kyo Inoue.
Generation of 1.5-$\mu$m band time-bin entanglement using spontaneous fiber four-wave mixing and planar light-wave circuit interferometers.
\emph{Physical Review A}, 72(4):041804, 2005.

\bibitem[Harder et~al.(2013)Harder, Ansari, Brecht, Dirmeier, Marquardt, and Silberhorn]{harder2013optimized}
Georg Harder, Vahid Ansari, Benjamin Brecht, Thomas Dirmeier, Christoph Marquardt, and Christine Silberhorn.
An optimized photon pair source for quantum circuits.
\emph{Optics Express}, 21(12):13975--13985, 2013.

\bibitem[Davanco et~al.(2012)Davanco, Ong, Shehata, Tosi, Agha, Assefa, Xia, Green, Mookherjea, and Srinivasan]{davanco2012telecommunications}
Marcelo Davanco, Jun Rong Ong, Andrea Bahgat Shehata, Alberto Tosi, Imad Agha, Solomon Assefa, Fengnian Xia, William M.~J. Green, Shayan Mookherjea, and Kartik Srinivasan.
Telecommunications-band heralded single photons from a silicon nanophotonic chip.
\emph{Applied Physics Letters}, 100(26), 2012.

\bibitem[Cordier et~al.(2020)Cordier, Delaye, G{\'e}r{\^o}me, Benabid, and Zaquine]{cordier2020raman}
Martin Cordier, Philippe Delaye, Fr{\'e}d{\'e}ric G{\'e}r{\^o}me, Fetah Benabid, and Isabelle Zaquine.
Raman-free fibered photon-pair source.
\emph{Scientific Reports}, 10(1):1650, 2020.

\bibitem[Zhong et~al.(2012)Zhong, Wong, Restelli, and Bienfang]{zhong2012efficient}
Tian Zhong, Franco N.~C. Wong, Alessandro Restelli, and Joshua C.~Bienfang.
Efficient single-spatial-mode periodically-poled KTiOPO$_4$ waveguide source for high-dimensional entanglement-based quantum key distribution.
\emph{Optics Express}, 20(24):26868--26877, 2012.

\bibitem[Troisi et~al.(2026)Troisi, Pelet, Dalidet, Sauder, Alibart, Tanzilli, and Martin]{Troisi2026Sagnac}
Tess Troisi, Yoann Pelet, Romain Dalidet, Gregory Sauder, Olivier Alibart, S{\'e}bastien Tanzilli, and Anthony Martin.
High-brightness fiber-based Sagnac source of entangled photon pairs for multiplexed quantum networks.
\emph{arXiv preprint arXiv:2602.08863}, 2026.

\bibitem[McKay(1982)]{testthesis}
R.~S. McKay.
\emph{X-ray Crystallography}.
PhD thesis, Princeton University, 1982.

\bibitem[Zhang et~al.(2014)Zhang, Qiao, Sun, Shi, Huang, Li, and Yang]{Zhang:14}
Yaxin Zhang, Shen Qiao, Linlin Sun, Qi Wu Shi, Wanxia Huang, Ling Li, and Ziqiang Yang.
Photoinduced active terahertz metamaterials with nanostructured vanadium dioxide film deposited by sol-gel method.
\emph{Optics Express}, 22(9):11070--11078, 2014.
doi: 10.1364/OE.22.011070.

\bibitem[Optica Publishing Group(n.d.)]{OPTICA}
Optica Publishing Group.
\emph{Optica}.
\url{https://opg.optica.org}.

\bibitem[Forster et~al.(2007)Forster, Ramaswamy, Artaxo, Bernsten, Betts, Fahey, Haywood, Lean, Lowe, Myhre, Nganga, Prinn, Raga, Schulz, and Dorland]{FORSTER2007}
P.~Forster, V.~Ramaswamy, P.~Artaxo, T.~Bernsten, R.~Betts, D.~Fahey, J.~Haywood, J.~Lean, D.~Lowe, G.~Myhre, J.~Nganga, R.~Prinn, G.~Raga, M.~Schulz, and R.~V. Dorland.
Changes in atmospheric constituents and in radiative forcing.
In \emph{Climate Change 2007: The Physical Science Basis. Contribution of Working Group 1 to the Fourth Assessment Report of the Intergovernmental Panel on Climate Change}.
Cambridge University Press, 2007.

\bibitem[Yelin et~al.(2003)Yelin, Oron, Thiberge, Moses, and Silberberg]{Yelin:03}
Dvir Yelin, Dan Oron, Stephan Thiberge, Elisha Moses, and Yaron Silberberg.
Multiphoton plasmon-resonance microscopy.
\emph{Optics Express}, 11(12):1385--1391, 2003.
doi: 10.1364/OE.11.001385.

\bibitem[Masajada et~al.(2013)Masajada, Bacia, and Drobczy\'nski]{Masajada:13}
Jan Masajada, Marcin Bacia, and S{\l}awomir Drobczy\'nski.
Cluster formation in ferrofluids induced by holographic optical tweezers.
\emph{Optics Letters}, 38(19):3910--3913, 2013.
doi: 10.1364/OL.38.003910.

\bibitem[Dean et~al.(2006)Dean, Aronstein, Smith, Shiri, and Acton]{Dean2006}
Bruce H.~Dean, David L.~Aronstein, Scott J.~Smith, Ron Shiri, and Scott D.~Acton.
Phase retrieval algorithm for JWST flight and testbed telescope.
In \emph{Space Telescopes and Instrumentation I: Optical, Infrared, and Millimeter}, p.~17, 2006.

\bibitem[Rivers(n.d.)]{codeexample}
C.~Rivers.
Epipy: Python tools for epidemiology.
\emph{figshare}, 2014.
Retrieved 13 May 2015.
\url{http://dx.doi.org/10.6084/m9.figshare.1005064}.

\end{thebibliography}




% \end{thebibliography}

\end{document}